\documentclass[aps,pra,twocolumn,superscriptaddress]{revtex4-2}
\usepackage{color}
\usepackage{array}
\usepackage{textcomp}
\usepackage{bm}
\usepackage{braket}
\usepackage{amsmath}
\usepackage{amssymb}
\usepackage{upgreek}
\usepackage{multirow}
\usepackage{graphicx}
\usepackage{hyperref}
\hypersetup{colorlinks=true, citecolor=blue, urlcolor=blue, linkcolor=blue}

\begin{document}
\title{Echoes in Unidirectionally Rotating Molecules}
\author{Long Xu}
\thanks{L. X., I. T., and L. Z. contributed equally to this work.}
\affiliation{AMOS and Department of Chemical and Biological Physics, Weizmann Institute
of Science, Rehovot 7610001, Israel}
\author{Ilia Tutunnikov}
\thanks{L. X., I. T., and L. Z. contributed equally to this work.}
\affiliation{AMOS and Department of Chemical and Biological Physics, Weizmann Institute
of Science, Rehovot 7610001, Israel}
\author{Lianrong Zhou}
\thanks{L. X., I. T., and L. Z. contributed equally to this work.}
\affiliation{State Key Laboratory of Precision Spectroscopy, East China Normal
University, Shanghai 200062, China}
\author{Kang Lin}
\affiliation{State Key Laboratory of Precision Spectroscopy, East China Normal
University, Shanghai 200062, China}
\author{Junjie Qiang}
\affiliation{State Key Laboratory of Precision Spectroscopy, East China Normal
University, Shanghai 200062, China}
\author{Peifen Lu}
\affiliation{State Key Laboratory of Precision Spectroscopy, East China Normal
University, Shanghai 200062, China}
\author{Yehiam Prior}
\thanks{yehiam.prior@weizmann.ac.il}
\affiliation{AMOS and Department of Chemical and Biological Physics, Weizmann Institute
of Science, Rehovot 7610001, Israel}
\affiliation{State Key Laboratory of Precision Spectroscopy, East China Normal
University, Shanghai 200062, China}
\author{Ilya Sh. Averbukh}
\thanks{ilya.averbukh@weizmann.ac.il}
\affiliation{AMOS and Department of Chemical and Biological Physics, Weizmann Institute
of Science, Rehovot 7610001, Israel}
\author{Jian Wu}
\thanks{jwu@phy.ecnu.edu.cn}
\affiliation{State Key Laboratory of Precision Spectroscopy, East China Normal
University, Shanghai 200062, China}
\affiliation{Collaborative Innovation Center of Extreme Optics, Shanxi University,
Taiyuan, Shanxi 030006, China}
\begin{abstract}
We report the experimental observation of molecular unidirectional
rotation (UDR) echoes, and analyze their origin and behavior both
classically and quantum mechanically. The molecules are excited by
two time-delayed polarization-twisted ultrashort laser pulses and
the echoes are measured by exploding the molecules and reconstructing
their spatial orientation from the detected recoil ions momenta. Unlike
alignment echoes which are induced by linearly polarized pulses, here
the axial symmetry is broken by the twisted polarization, giving rise
to molecular unidirectional rotation. We find that the rotation sense
of the echo is governed by the twisting sense of the second pulse
even when its intensity is much weaker than the intensity of the first
pulse. In our theoretical study, we rely on classical phase space
analysis and on three-dimensional quantum simulations of the laser-driven
molecular dynamics. Both approaches nicely reproduce the experimental
results. Echoes in general, and the unique UDR echoes in particular,
provide new tools for studies of relaxation processes in dense molecular
gases.
\end{abstract}
\maketitle

\section{Introduction \label{sec:introduction}}

In 1950, E. L. Hahn discovered \citep{Hahn1950,Hahn1953} that if
an ensemble of spins is irradiated by two properly timed and shaped
magnetic field pulses, a spontaneous magnetization of the sample appears
at twice the time delay between the two pulses. This response was
termed ``spin echo''. This original discovery is a fundamental element
in modern Nuclear Magnetic Resonance (NMR) and Magnetic Resonance
Imaging (MRI). Following the spin echoes, echoes have been observed
in various non-linear physical systems, including photon echoes \citep{Kurnit1964,Mukamel1995},
cyclotron echoes \citep{Hill1965}, plasma-wave echoes \citep{Gould1967},
cold atom echoes in optical traps \citep{Bulatov1998,Buchkremer2000,Herrera2012},
and echoes in particle accelerators \citep{Stupakov1992,Stupakov1993,Spentzouris1996,Stupakov2013,Sen2018}.
The concept of echo was extended to single quantum particles and the
phenomenon was observed in atoms coupled to a single-mode cavity \citep{Meunier2005}
and in a single vibrationally excited molecule \citep{Qiang2020}.

In the last five years, a new class of molecular alignment and orientation
echoes has been discovered and extensively investigated \citep{Karras2015Orientation,Karras2016Experimental,Lin2016,Lin2017,Rosenberg2017,Wang2019,Lin2020}.
Briefly, a non-resonant optical pulse polarizes the molecules and
interacts with the induced dipole, resulting in a torque proportional
to $\mathbf{d}_{\mathrm{ind}}\times\mathbf{E}$, where $\mathbf{d}_{\mathrm{ind}}$
is the dipole moment induced by the electric field $\mathbf{E}$ of
the pulse. When a pulse is short compared to the typical molecular
rotational period, it impulsively kicks the molecules such that shortly
after the pulse, the molecules are aligned along a direction defined
by the field polarization (for reviews of molecular alignment see
Refs. \citep{Stapelfeldt2003,Ohshima2010,Fleischer2012,Lemeshko2013,Koch2019}).
With time, molecules with different angular velocities step out of
phase and the alignment disappears. However, in a manner similar to
the original spin echoes, it was shown (see e.g. \citep{Karras2015Orientation,Lin2016})
that if the molecules are kicked again by a second, delayed laser
pulse, a spontaneous alignment response emerges at twice the delay
between the pulses. Echoes of this type have been used in studies
of molecular relaxation in condensed gases \citep{Rosenberg2018,Zhang2019,Ma2019,Rosenberg2020}.

Molecules kicked by linearly polarized pulses are equally likely to
rotate clockwise or counterclockwise, hence after averaging all molecules
there is no preferred sense of molecular rotation and the ensemble-averaged
angular momentum remains zero. Besides linearly polarized pulses,
polarization shaped pulses have also been used to drive molecular
rotation. These include pulses in which the polarization of the electric
laser field rotates in a plane during the pulse, termed polarization-twisted
pulses. Examples of such pulses include a pair of delayed cross-polarized
laser pulses \citep{Fleischer2009,Kitano2009,Khodorkovsky2011}, a
chiral train of ultrashort pulses \citep{Zhdanovich2011,Bloomquist2012},
polarization shaped pulses \citep{Kida2008,Kida2009,Karras2015Polarization,Prost2017,Mizuse2020},
and the optical centrifuge for molecules \citep{Karczmarek1999,Villeneuve2000,Yuan2011,Korobenko2014,Korobenko2018}.
While these pulses are all different, in all cases the field polarization
twists in a plane perpendicular to the propagation direction, and
the time scale of the polarization twisting is comparable to that
of the molecular rotation. This allows the molecules to follow the
twisting polarization, resulting in \emph{molecular unidirectional
rotation} (UDR), and a non-zero ensemble-averaged angular momentum
perpendicular to the plane of polarization twisting. It is not surprising
that the dynamics induced by polarization-twisted pulses is different
from the alignment dynamics induced by linearly polarized pulses.

In this paper, we report the experimental observation of molecular
UDR echo of linear molecules excited by polarization-twisted pulses,
and provide a detailed theoretical analysis of this effect. In our
experiments, the UDR echoes are induced by a pair of co-rotating or
counter-rotating delayed polarization-twisted pulses. Each pulse induces
molecular UDR, i.e. molecules come to alignment and while being aligned
rotate together for a short period of time. At twice the delay between
the pulses, molecular UDR spontaneously reoccur. This recurrence is
what we call UDR echo.

One of the techniques for creating polarization-twisted pulse is by
partially temporally overlapping two delayed orthogonally polarized
femtosecond laser pulses. The polarization of the resulting laser
field rotates continuously from the direction of polarization of the
first pulse to that of the second one, thus performing a quarter rotation
in a plane \citep{Kida2008,Kida2009,Karras2015Polarization,Prost2017}.
The molecular dynamics is followed by using the detection methodology
of COLTRIMS (cold target recoil ion momentum spectroscopy) \citep{DORNER2000Cold,Ullrich2003}.
In this technique, an intense probe pulse comes at a variable delay
after the pump pulses and Coulomb-explodes the molecules. The charged
molecular fragments are then guided to a coincidence ion detector
allowing the recording of the three-dimensional (3D) momenta of the
particles. Based on this information, the molecular angular distribution
at the moment of the Coulomb explosion can be reconstructed.

The present work introduces several new elements beyond previous works
on molecular alignment/orientation echoes. First, in the case of excitation
by polarization-twisted pulses, the system lacks axial symmetry and
its full characterization involves two degrees of freedom---polar
and azimuthal angles. The COLTRIMS detection scheme provides the full
angular information and allows to observe the echo response in both
degrees of freedom. Second, the echo observed here is of a new, previously
unreported, type. During the echo, the molecules not only come to
alignment, but the alignment direction rotates with a preferred sense.
Perhaps the most surprising observation is that the second polarization-twisted
pulse controls the sense of the rotation of the echo. This is unexpected,
since the second pulse is less intense and does not change the sign
of the ensemble-averaged angular momentum defined by the first polarization-twisted
pulse.

The paper is organized as follows. In Sec. \ref{sec:Phase-space-analysis}
we consider a simplified classical model in which the molecular rotation
is restricted to a plane, and we discuss the UDR echo from several
points of view. Sections \ref{sec:Experimental-Methods} and \ref{sec:Numerical-methods}
summarize our experimental and numerical methods, respectively. In
Sec. \ref{sec:Experimental-results} we present the experimental results
and compare them with the results of three-dimensional classical and
quantum simulations. Finally, the discussion in Sec. \ref{sec:Conclusions}
summarizes the paper.

\section{CLASSICAL PERSPECTIVE \label{sec:Phase-space-analysis}}

\subsection{Introduction to molecular UDR echo}

Consider a simplified classical model, in which the molecules are
rigid, and do not interact with each other. The laser pulses propagate
along the $X$ axis, the molecules are restricted to rotate in the
$YZ$ plane, and the interaction potential due to the laser electric
field (averaged over the optical cycle) is given by $V(\psi,t)=-(\Delta\alpha/4)E^{2}(t)\cos^{2}(\psi)$
\citep{boyd1992nonlinear,Friedrich1995Alignment,Friedrich1995Polarization}.
Here $\psi$ is the angle between the molecular axis and the field
polarization direction, $\Delta\alpha$ is the molecular polarizability
anisotropy, and $E(t)$ is the slowly varying envelope of the laser
pulse. We define an additional angle $\phi$, the angle between the
molecular and $Y$ axes. The equation of motion for $\phi$ is given
by
\begin{align}
\frac{\mathrm{d}^{2}\phi(t)}{\mathrm{d}t^{2}}=-\frac{1}{I}\frac{\mathrm{d}V}{\mathrm{d}\psi}=-\frac{\Delta\alpha E^{2}(t)}{4I}\sin\left[2\phi(t)-2\chi(t)\right],\label{eq:2Dmodel}
\end{align}
where $I$ is the moment of inertia of the molecule, $\chi=\phi-\psi$
is the angle between the field polarization direction and the $Y$
axis.

\begin{figure}[!b]
\centering{}\includegraphics[width=0.6\columnwidth]{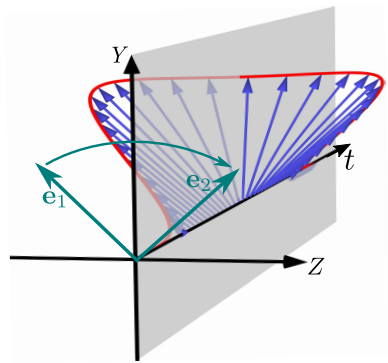} \caption{Illustration of the electric field of polarization-twisted laser pulse.
Here, the polarization rotates from the direction defined by $\mathbf{e}_{1}$
towards the direction of $\mathbf{e}_{2}$ in a clockwise sense. Both
of the vectors $\mathbf{e}_{1}$ and $\mathbf{e}_{2}$ are at angle
$\pi/4$ with respect to the $Y$ axis. \label{fig:pulse_illustration}}
\end{figure}

To simulate the behavior of a molecular ensemble, we use the Monte
Carlo approach and consider a collection of $N$ molecules. Equation
(\ref{eq:2Dmodel}) is solved numerically for each molecule. Initially,
the angle $\phi$ is uniformly distributed in the interval $(-\pi,\pi]$,
while the distribution of angular velocity $\omega\equiv\mathrm{d}\phi/\mathrm{d}t$,
is given by the Boltzmann distribution $p(\omega_{0})\propto\exp[-I\omega_{0}^{2}/(2k_{B}T)]$,
where $T$ is temperature and $k_{B}$ is the Boltzmann constant.
In the present work, both in theoretical analysis and in the experiment,
we use $\mathrm{N_{2}O}$ molecules and set the initial rotational
temperature to $T=75$ K. The molecular parameters were computed with
the help of the density functional theory (method: CAM-B3LYPultrafine/aug-ccpVTZ),
$I=39.786$ $\mathrm{amu}\cdot\mathrm{\mathring{A}}^{2}$ and $\Delta\alpha=2.794$
$\mathrm{\mathring{A}}^{3}$ \citep{johnson1999nist}.

A polarization-twisted pulse is modeled as a pair of overlapping orthogonally
polarized laser pulses with a delay $\tau_{p}$ between them
\begin{align}
\mathbf{E}(t)=E_{0}\left[f(t)\mathbf{e}_{1}+f(t-\tau_{p})\mathbf{e}_{2}\right],\label{eq:TwistedPulse}
\end{align}
where $E_{0}$ is the electric field peak amplitude, $\mathbf{e}_{1}$
and $\mathbf{e}_{2}$ are the unit vectors defining the polarizations
of the two pulses and are directed along $\pi/4$ (second quadrant)
and $-\pi/4$ (first quadrant) with respect to the $Y$ axis, respectively
(see Fig. \ref{fig:pulse_illustration}). The envelope of each constituent
linearly polarized pulse is given by $f(t)=\exp[-2\ln2(t/\sigma)^{2}]$,
where $\sigma$ is the full width at half maximum of the laser pulse
\emph{intensity }profile.

\begin{figure}[!b]
\begin{centering}
\includegraphics[width=7cm]{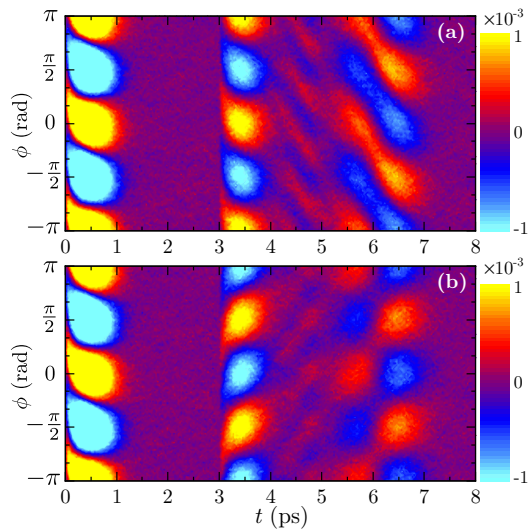}
\par\end{centering}
\caption{Time-dependent angular distributions for the two cases, (a) both polarization-twisted
pulses $\mathrm{P_{1}}$ and $\mathrm{P_{2}}$ twist in the same clockwise
sense ($\mathbf{e}_{1}\rightarrow\mathbf{e}_{2}$, see Fig. \ref{fig:pulse_illustration}),
and (b) $\mathrm{P_{1}}$ twists clockwise, while $\mathrm{P_{2}}$
twists counterclockwise ($\mathbf{e}_{1}\rightarrow-\mathbf{e}_{2}$).
Pulses' parameters: $\sigma=38$ fs, $\tau_{p}=48$ fs {[}see Eq.
(\ref{eq:TwistedPulse}){]}. Peak intensities of $\mathrm{P_{1}}$
and $\mathrm{P_{2}}$ are $5.0\times\;10^{13}\;\mathrm{W/cm^{2}}$
and $1.9\times\;10^{13}\;\mathrm{W/cm^{2}}$, respectively. \label{fig:angular_distributions_echo}}
\end{figure}

Figure \ref{fig:angular_distributions_echo} shows the time-dependent
angular distribution of a molecular ensemble excited by a pair of
time-delayed polarization-twisted laser pulses. The first pulse $\mathrm{P_{1}}$
is applied at $t=0$, and the second, less intense, pulse $\mathrm{P_{2}}$
is applied at $t=3$ ps. For better visibility, we subtract the averaged,
between $t=1.5$ ps and $t=2.5$ ps, angular distribution. In the
case of Fig. \ref{fig:angular_distributions_echo}(a), both pulses
twist in the same clockwise sense ($\mathbf{e}_{1}\rightarrow\mathbf{e}_{2}$,
see Fig. \ref{fig:pulse_illustration}). During the interaction with
both pulses, the molecules primarily align along the $Y$ axis ($\phi=0$),
while the shape of the distribution suggests that the molecules follow
the polarization twisting and rotate clockwise. The contributions
to the ensemble-averaged angular momentum of the two pulses are accumulated
in this case, because $\mathrm{P_{1}}$ and $\mathrm{P_{2}}$ twist
in the same sense. The echo response emerges at twice the delay between
$\mathrm{P_{1}}$ and $\mathrm{P_{2}}$ (at $t\approx6$ ps), which
is a general feature of the echo phenomenon (see the references in
Sec. \ref{sec:introduction}). During the echo, molecules transiently
align at $\phi=0,\pi$ and while remaining aligned, rotate towards
$\phi=\pm\pi/2$. Then, due to the dispersion of molecular angular
velocities, the alignment disappears, and with it the echo response.
After an additional delay of 3 ps (at $t\approx9$ ps), a 2nd order
echo \citep{Stupakov2013,Karras2015Orientation,Karras2016Experimental,Lin2016,Mizuse2020}
can be observed (not shown here), and a trace of a 3rd order echo
can be identified at $t\approx12$ ps (not shown here).

Figure \ref{fig:angular_distributions_echo}(b) shows the echo response
for a case when $\mathrm{P_{1}}$ and $\mathrm{P_{2}}$ twist in opposite
senses: $\mathrm{P}_{1}$ twists clockwise ($\mathbf{e}_{1}\rightarrow\mathbf{e}_{2}$,
see Fig. \ref{fig:pulse_illustration}) inducing a negative ensemble-averaged
angular momentum, while $\mathrm{P}_{2}$ twists counterclockwise
($\mathbf{e}_{1}\rightarrow-\mathbf{e}_{2}$, see Fig. \ref{fig:pulse_illustration})
inducing a positive ensemble-averaged angular momentum. In contrast
to the case of co-rotating pulses {[}Fig. \ref{fig:angular_distributions_echo}(a){]},
the contributions of the two pulses to the ensemble-averaged angular
momentum are of opposite signs. After the second pulse, the angular
momentum remains negative, because the second pulse is not intense
enough to cancel the angular momentum introduced by the first pulse.
Interestingly, the echo emerging at $t\approx6$ ps rotates counterclockwise,
which is \emph{against} the ensemble-averaged rotation sense of the
ensemble (clockwise sense, corresponding to the on average negative
angular velocity/momentum, see Appendix \ref{App:Population difference}).
In the following subsections, we discuss the echo formation mechanism
as well as the rotation sense of the molecules forming the echo.

\subsection{Echo formation: phase space analysis}

\begin{figure}[b]
\begin{centering}
\includegraphics[width=1\columnwidth]{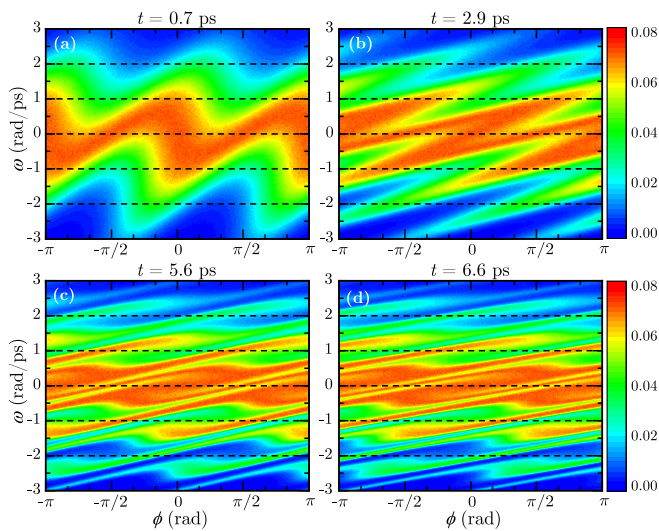}
\par\end{centering}
\caption{Phase space dynamics. Here, the polarization-twisted pulses $\mathrm{P}_{1}$
and $\mathrm{P_{2}}$ twist in the same clockwise sense ($\mathbf{e}_{1}\rightarrow\mathbf{e}_{2}$,
see Fig. \ref{fig:pulse_illustration}) and are applied at $t=0$
and $t=3$ ps, respectively. The parameters are the same as for the
case shown in Fig. \ref{fig:angular_distributions_echo}(a). (a) Shortly
after the first pulse. The phase space distribution shows transient
alignment along the $Y$ axis (higher probability near $\phi=0,\pm\pi$).
(b) Filamented phase space just before the second pulse. (c) The moment
of alignment echo, higher probability near $\phi=0,\pm\pi$, just
like after the first pulse. (d) The aligned molecules rotate unidirectionally
resulting in antialignment phase of the echo (higher probability near
$\phi=\pm\pi/2$). \label{fig:Phase_Space}}
\end{figure}

Following earlier works \citep{Stupakov1993,Karras2015Orientation,Karras2016Experimental,Lin2016},
we discuss the echo formation mechanism from the geometrical point
of view by considering the phase space dynamics. Figure \ref{fig:Phase_Space}
shows the phase space distribution, $P(\phi,\omega)$ at different
times for the case when both polarization-twisted pulses $\mathrm{P_{1}}$
and $\mathrm{P_{2}}$ twist in the same sense {[}as depicted in Fig.
\ref{fig:angular_distributions_echo}(a){]}. Figure \ref{fig:Phase_Space}(a)
shows that shortly after the first pulse $\mathrm{P_{1}}$, the molecules
come to alignment along the $Y$ axis {[}higher probability (red)
around $\phi=0,\pi${]}. In this sense, a short polarization-twisted
pulse resembles a pulse with a fixed linear polarization along the
$Y$ axis. With time, the phase space distribution develops multiple
parallel filaments {[}see Fig. \ref{fig:Phase_Space}(b){]}. As the
number of the filaments grow, they become thinner due to the conservation
of the phase space volume, and they form a ``quasi-quantized'' phase
space pattern \citep{Karras2015Orientation,Karras2016Experimental,Lin2016}.
Such filamentation of phase space is a known phenomenon in stellar
systems \citep{Lynden1967Statistical} and in accelerator physics
\citep{lichtenberg1969phase}. In previous works \citep{Karras2015Orientation,Karras2016Experimental,Lin2016},
it was shown that the phase space filamentation induced by a single
linearly polarized pulse is symmetric with respect to the phase space
origin $(\phi,\omega)=(0,0)$. As a result, the ensemble-averaged
angular velocity/momentum vanishes. Here, in contrast, the polarization-twisted
pulse initiates molecular UDR, breaking this symmetry. The total probabilities
of finding molecules with positive/negative angular velocity (above/below
the horizontal line $\omega=0$) are unequal. In other words, the
ensemble-averaged angular velocity (momentum) is different from zero.

After the interaction with the second polarization-twisted pulse $\mathrm{P_{2}}$
(applied at $t=3$ ps), every filament shown in Fig. \ref{fig:Phase_Space}(b)
folds in a manner similar to Fig. \ref{fig:Phase_Space}(a), including
the aforementioned asymmetry with respect to the origin. Then, after
an additional delay of 3 ps, at $t\approx6$ ps, the folded parts
of the filaments pile-up at $\phi=0,\pi$ {[}see Fig. \ref{fig:Phase_Space}(c){]},
representing spontaneous molecular alignment along the $Y$ axis.
This recurrence (echo) of the molecular alignment happens due to the
``quasi-quantization'' of the phase space distribution, as shown
in Fig. \ref{fig:Phase_Space}(b). The aligned molecules rotate unidirectionally,
such that the probability increases near $\phi=\pm\pi/2$ {[}see Fig.
\ref{fig:Phase_Space}(d){]}.

\begin{figure}[!b]
\begin{centering}
\includegraphics[width=1\columnwidth]{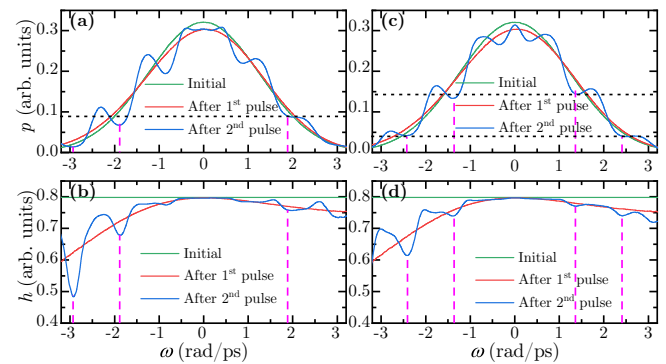}
\par\end{centering}
\caption{(a) The distribution of angular velocity, $p(\omega)$ and (b) the
corresponding information entropy, $h(\omega)$ for the case shown
in Fig. \ref{fig:angular_distributions_echo}(a), where both pulses
twist clockwise from $\pi/4$ to $-\pi/4$. (c,d) The same as (a,b),
except that the second pulse twists counterclockwise from $\pi/4$
to $3\pi/4$. \label{fig:Entropy}}
\end{figure}

\subsection{Echo formation: angular velocity distribution}

In this subsection, we take a closer look at the distribution of the
angular velocity, $p(\omega)$ at different times. $p(\omega)$ is
the marginal distribution obtained by integrating the phase space
distribution with respect to the angle $\phi$, $p(\omega)=\int_{-\pi}^{\pi}P(\phi,\omega)\mathrm{d}\phi$.
Figure \ref{fig:Entropy}(a) shows the distributions, $p(\omega)$
before the application of polarization-twisted pulses, after a single
pulse $\mathrm{P}_{1}$, and after both pulses $\mathrm{P}_{1}$ and
$\mathrm{P}_{2}$ are applied. Here, $\mathrm{P}_{1}$ and $\mathrm{P}_{2}$
twist in the same clockwise sense ($\mathbf{e}_{1}\rightarrow\mathbf{e}_{2}$,
see Fig. \ref{fig:pulse_illustration}), similar to Fig. \ref{fig:angular_distributions_echo}(a)
and Fig. \ref{fig:Phase_Space}. Obviously, the fractions of molecules
rotating clockwise and counterclockwise are equal before the pulses.
After $\mathrm{P}_{1}$ (twisting clockwise), there are more molecules
rotating clockwise than counterclockwise ($\int_{0}^{\infty}p(\omega)\mathrm{d}\omega<\int_{-\infty}^{0}p(\omega)\mathrm{d}\omega$,
see Appendix \ref{App:Population difference}). Furthermore, since
$\mathrm{P}_{2}$ twists in the same clockwise sense, this imbalance
is enhanced after the application of the second pulse. It is evident
from Fig. \ref{fig:Entropy}(a) that $p(\omega)$ is a slightly asymmetric
function of $\omega$. Accordingly, the average angular velocity,
$\overline{\omega}=\int\omega p(\omega)\mathrm{d}\omega$ vanishes
before the pulses, and is negative after a single pulse $\mathrm{P}_{1}$
and after the application of both pulses (see Appendix \ref{App:Population difference}).
In the case of linearly polarized pulses, the distribution $p(\omega)$
always remains symmetric, and $\overline{\omega}=0$. To identify
the subgroups of molecules contributing to the echo formation, we
use the information entropy \citep{Shannon1948A,cover1991elements}
defined by
\begin{equation}
h(\omega)=-\int_{-\pi}^{\pi}g(\phi,\omega)\log\left[g(\phi,\omega)\right]\mathrm{d}\phi,
\end{equation}
where $g(\phi,\omega)\mathrm{d}\phi=P(\phi,\omega)\mathrm{d}\phi/p(\omega),$
($\int_{-\pi}^{\pi}g(\phi,\omega)\mathrm{d}\phi=1$) is the probability
of finding a molecule in a small angular interval $\mathrm{d}\phi$
around the angle $\phi$ at angular velocity $\omega$. The information
entropy $h(\omega)$ quantifies the difference between a given distribution
and a uniform distribution.

Figure \ref{fig:Entropy}(b) shows the information entropy for the
three distributions presented in Fig. \ref{fig:Entropy}(a). Note
that the angular velocity remains constant during the stages of free
evolution (in the absence of external fields). The initial information
entropy is constant (and maximum), as expected for the initially uniform
angular distribution. After $\mathrm{P}_{1}$, the entropy $h(\omega)$
is lower for negative angular velocities. This means that $\mathrm{P}_{1}$
(twisting clockwise $\mathbf{e}_{1}\rightarrow\mathbf{e}_{2}$, see
Fig. \ref{fig:pulse_illustration}) not only induces molecular UDR,
but also increases the angular order of molecules with negative angular
velocity. The second polarization-twisted pulse $\mathrm{P}_{2}$
twists in the same clockwise sense and modulates the distribution
of angular velocities {[}see Fig. \ref{fig:Entropy}(a), blue curve{]},
which is reflected in the information entropy. Figure \ref{fig:Entropy}(b)
shows that after $\mathrm{P}_{2}$, the information entropy develops
a series of minima located at $\omega\approx\pm0.838,\pm1.885,\pm2.932\,\mathrm{rad/ps}$
corresponding to subgroups of ordered molecules (their angular distribution
differs from the uniform distribution) moving with the characteristic
angular velocities.

The contribution to the alignment echo comes mainly from these groups
because uniformly distributed molecules do not contribute to the alignment.
The difference of angular velocities of the adjacent ordered groups
is approximately $1.047\,\mathrm{rad/ps}$, leading to the resynchronization
of these groups after $(\pi\,\mathrm{rad})/(1.047\,\mathrm{rad/ps})\approx3$
ps following $\mathrm{P}_{2}$, in agreement with Fig. \ref{fig:angular_distributions_echo}
and Fig. \ref{fig:Phase_Space}.

As an additional example, we consider the case when the second pulse
$\mathrm{P}_{2}$ twists in the opposite sense (counterclockwise $\mathbf{e}_{1}\rightarrow-\mathbf{e}_{2}$,
see Fig. \ref{fig:pulse_illustration}). Figure \ref{fig:Entropy}(c)
shows the resulting angular velocity distribution, while Fig. \ref{fig:Entropy}(d)
shows the corresponding information entropy. In this case, $\mathrm{P}_{2}$
competes with $\mathrm{P}_{1}$ and tends to rotate the molecules
in the opposite direction. Here, $\mathrm{P}_{1}$ has higher intensity,
such that on average (including all molecules), the molecules continue
rotating clockwise (after both pulses, $\overline{\omega}<0$, see
Appendix \ref{App:Population difference}). The positions of the minima
of information entropy in this case are $\omega\approx\pm0.314,\pm1.361,\pm2.408\,\mathrm{rad/ps}$
{[}see Fig. \ref{fig:Entropy}(d){]} and are shifted as compared to
the minima in the case when both pulses twist in the same sense {[}$\omega\approx\pm0.838,\pm1.885,\pm2.932\,\mathrm{rad/ps}$,
see Fig. \ref{fig:Entropy}(b){]}. Besides the shift, the difference
of angular velocities between the adjacent minima is the same, $1.047\,\mathrm{rad/ps}$,
leading to the echo response after $(\pi\,\mathrm{rad})/(1.047\,\mathrm{rad/ps})\approx3\,\mathrm{ps}$,
like in the first case {[}see Fig. \ref{fig:angular_distributions_echo}(b){]}.

\subsection{Effects of the two pulses on the Molecular UDR echo}

\begin{figure}[!bthp]
\begin{centering}
\includegraphics[width=1\columnwidth]{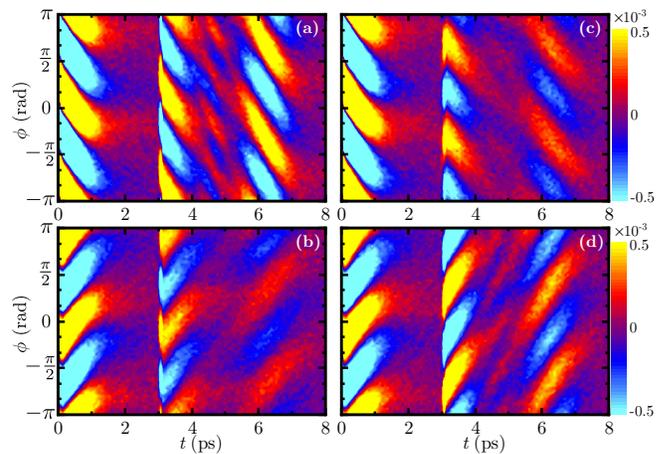}
\par\end{centering}
\caption{On the left: Time-dependent angular distributions contributed by the
molecules with (a) negative and (b) positive angular velocities for
the case shown in Fig. \ref{fig:angular_distributions_echo}(a). On
the right: Time-dependent angular distributions contributed by the
molecules with (c) negative and (d) positive angular velocities for
the case shown in Fig. \ref{fig:angular_distributions_echo}(b). \label{fig:pos_neg_dist}}
\end{figure}

Figure \ref{fig:angular_distributions_echo} shows that the UDR echo
follows the twisting sense of the second polarization-twisted pulse
$\mathrm{P}_{2}$ irrespective of whether it twists in the same or
opposite sense relative to the first polarization-twisted pulse $\mathrm{P}_{1}$.
This is counter intuitive since $\mathrm{P}_{2}$ is less intense,
and on average, the molecules rotate in the sense determined by $\mathrm{P}_{1}$
(see Appendix \ref{App:Population difference}).

Figure \ref{fig:pos_neg_dist} shows the angular distributions of
molecules with negative and positive angular velocities separately
for the two cases shown in Fig. \ref{fig:angular_distributions_echo}.
It is evident from Figs. \ref{fig:pos_neg_dist}(a) and \ref{fig:pos_neg_dist}(c)
that molecules with negative angular velocities come to alignment
and collectively rotate clockwise during the echo at $t\approx6$
ps. Likewise, Figs. \ref{fig:pos_neg_dist}(b) and \ref{fig:pos_neg_dist}(d)
show that molecules with positive angular velocities come to alignment
and rotate counterclockwise during the echo. In addition, Fig. \ref{fig:pos_neg_dist}
suggests that the second polarization-twisted pulse $\mathrm{P}_{2}$
enhances the alignment and accelerates the molecules whose sense of
rotation matches the sense of polarization twisting of $\mathrm{P}_{2}$.
Having a relatively high alignment factor, these molecules determine
the behavior during the echo.

The described molecular UDR echoes are in sharp contrast to the molecular
alignment echoes induced by a sequence of linearly polarized pulses
\citep{Karras2015Orientation,Karras2016Experimental,Lin2016}. In
case of excitation by linearly polarized pulses, the distribution
of angular velocities always remains symmetric, such that during the
echo, the aligned molecules do not rotate with a preferred sense.

\section{Experimental and Numerical Methods}

\begin{figure}[b]
\begin{centering}
\includegraphics[width=1\columnwidth]{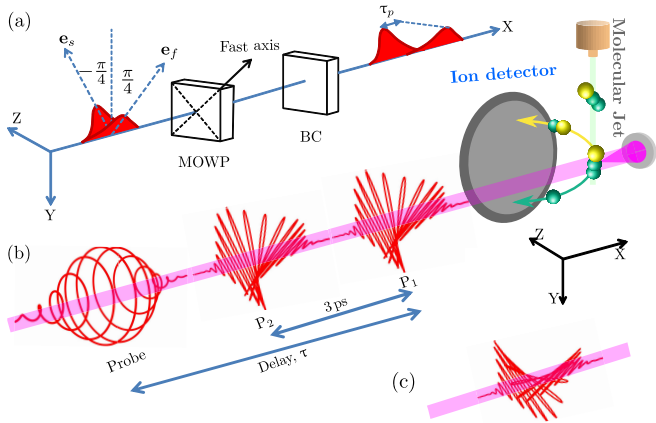}
\par\end{centering}
\caption{(a) Optical setup for generating polarization-twisted pulses (see
Ref. \citep{Karras2015Polarization}) described by Eq. (\ref{eq:ExpField}).
A linearly polarized (along the $Y$ axis) pulse propagates along
the $X$ axis and passes through a multiple order wave plate (MOWP)
whose fast axis is at $\pi/4$ with respect to the $Y$ axis introducing
a delay $\tau_{p}$ ($48\,\mathrm{fs}$ in our experiment) between
the pulse polarization projections (along $\mathbf{e}_{f}$ and $\mathbf{e}_{s}$).
The pulse continues through a Berek compensator (BC). The introduced
delay between the polarization components results in a pulse with
polarization twisting from the fast axis $\mathbf{e}_{f}$ towards
the slow axis $\mathbf{e}_{s}$. (b) A schematic diagram of the experimental
setup (see text). (c) Illustration of polarization-twisted pulse $\mathrm{P}_{2}$
used in the second experiment. \label{fig:setup}}
\end{figure}

\subsection{Experimental Setup \label{sec:Experimental-Methods}}

The measurements were performed in the ultrahigh vacuum chamber of
a COLTRIMS apparatus \citep{DORNER2000Cold,Ullrich2003}. A sequence
of laser pulses is focused (by a concave silver mirror with a focal
length of $f=75$ mm) on a supersonic gas jet of $\mathrm{N_{2}O}$
molecules. The laser pulses are produced by a multipass amplifier
Ti:sapphire laser system ($800\,\mathrm{nm},\;10\,\mathrm{kHz},\;25\,\mathrm{fs}$).
In our experiments, we induce molecular UDR using polarization-twisted
laser pulses. The pulses are generated following the polarization
shaping technique introduced in Ref. \citep{Karras2015Polarization}.
Briefly, a linearly polarized pulse passes through a multiple order
wave plate (MOWP) splitting it into two orthogonally polarized time-delayed
(but overlapping) pulses {[}see the setup in Fig. \ref{fig:setup}(a){]}.
The electric field of the resulting laser pulse can be described by
\citep{Karras2015Polarization}
\begin{align}
\mathbf{E} & =E_{s}\left(t-\tau_{p}\right)\cos\left(\omega_{1}t-\Delta\varphi\right)\mathbf{e}_{s},\nonumber \\
 & +E_{f}(t)\cos(\omega_{1}t)\mathbf{e}_{f}\label{eq:ExpField}
\end{align}
where $\omega_{1}$ is the central angular frequency, $\tau_{p}$
is the time shift between the crossed pulses and can be adjusted by
MOWP, $\mathbf{e}_{s}$ and $\mathbf{e}_{f}$ are the unit vectors
along the slow ($-\pi/4$ with respect to the $Y$ axis) and fast
($\pi/4$ with respect to the $Y$ axis) axes, respectively. Here,
the total relative phase $\Delta\varphi$ is defined by $\Delta\varphi=\omega_{1}\tau_{p}+\varphi_{p}$,
where $\varphi_{p}$ is the phase difference caused and controlled
by Berek compensator (BC). As depicted in Fig. \ref{fig:setup}(b),
a total of three femtosecond laser pulses are employed in the experiments:
polarization-twisted pulses $\mathrm{P}_{1}$, $\mathrm{P}_{2}$,
and the circularly polarized probe pulse. The pulse ${\rm P_{2}}$
follows ${\rm P_{1}}$ with a 3 ps delay and induces the molecular
UDR echo after an additional delay of 3 ps. The electric field of
both $\mathrm{P}_{1}$ and ${\rm P_{2}}$ is described by Eq. (\ref{eq:ExpField}),
where the value of $\Delta\varphi$ is set to zero in order to obtain
a linear polarization twisting in a plane. For $0<\Delta\varphi<\pi$,
the polarization in the overlap region is elliptical.

An intense circularly polarized probe pulse is sent at a variable
delay $\tau$, which Coulomb explodes the $\mathrm{N}_{2}\mathrm{O}$
molecules. The ionized fragments, accelerated by a static electric
field ($\approx22.3\,\mathrm{V}/\mathrm{cm}$), fly towards the position-sensitive
microchannel plate detector and time-sensitive delay line detector
at one end of the spectrometer. During the offline analysis, the instantaneous
$3\mathrm{D}$-momentum of the fragments during the Coulomb explosion
can be reconstructed based on the coincidence measurements. In the
experiments, the peak intensities of the pulses $\mathrm{P}_{1}$,
$\mathrm{P}_{2}$, and probe pulses are approximately $5.0\times10^{13}\,\mathrm{W}/\mathrm{cm}^{2}$,
$1.9\times10^{13}\,\mathrm{W}/\mathrm{cm}^{2}$, and $4.0\times10^{14}\,\mathrm{W}/\mathrm{cm}^{2}$,
respectively. The temporal duration of both $\mathrm{P}_{1}$ and
$\mathrm{P}_{2}$ pulses is approximately 38 fs {[}see Eq. (\ref{eq:ExpField}){]}.
To explore the time-dependent evolution of molecular dynamics, the
probe delay (measured from $\mathrm{P}_{1}$ ) is scanned from $0$
to $49$ ps. Since the molecular rotational period is much longer
than the time scale of the Coulomb explosion process, the orientation
of the molecular axis can be recovered from the reconstructed 3D-momentum
vectors of the ions. Based on the experimentally measured translation
temperature of the molecules \citep{Lin2016}, the upper limit of
the rotational temperature of $\mathrm{N}_{2}\mathrm{O}$ molecules
in our experiments is estimated to be $\approx75\,\mathrm{K}$.

\subsection{Numerical methods \label{sec:Numerical-methods}}

In what follows, we simulate the echo dynamics under the experimental
conditions. For this, we model the laser-driven molecular rotation
in three dimensions. Both classical and quantum mechanical formalisms
are used, and the results are compared to the experiments.\\

\noindent \textbf{Classical simulation.} An efficient strategy for
simulating the rotational dynamics of an ensemble of rigid bodies
relies on the Monte Carlo approach. A sample consisting of $N$ molecules
is propagated in time, and the appropriate observables are obtained
by evaluating the averages over the sample. Generally, the motion
of each member of the sample is obtained by solving the Euler's equations
\citep{Goldstein2002Classical}. The absolute orientation in space
is parameterized by quaternions \citep{Coutsias2004The,Kuipers1999Quaternions},
which are generalized complex numbers having four components. This
approach was used in our previous works \citep{Tutunnikov2019Laser,xu2020longlasting},
where the full description of the method is provided. Here, we apply
the scheme to the linear molecules and model the electric field of
the laser pulses by Eq. (\ref{eq:TwistedPulse}).\\

\begin{figure}[!b]
\begin{centering}
\includegraphics[width=1\columnwidth]{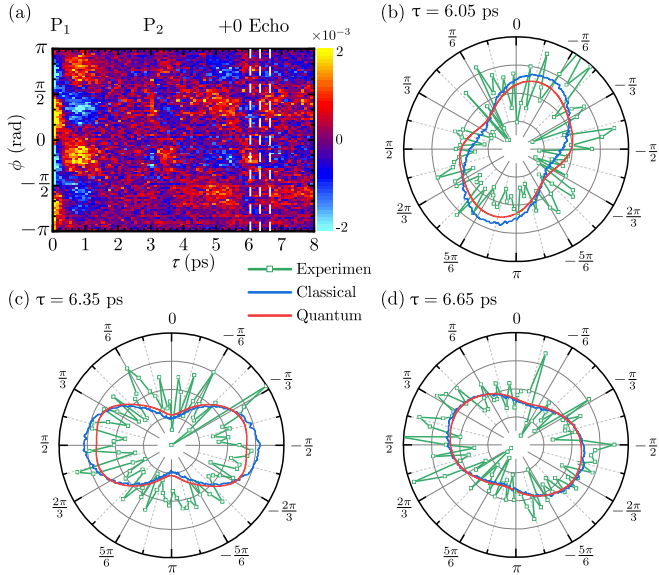}
\par\end{centering}
\caption{(a) Experimentally measured time-dependent angular distribution as
a function of time for the case when both polarization-twisted pulses
$\mathrm{P_{1,2}}$ twist in a clockwise sense from $\pi/4$ to $-\pi/4$.
(b-d) Polar plots of the experimentally measured (green lines), simulated
classically (blue lines), and quantum mechanically (red lines) molecular
angular distribution during the UDR echo at (b) $\tau=6.05$ ps, (c)
$\tau=6.35$ ps, and (d) $\tau=6.65$ ps. \label{fig:Experiment_condition1}}
\end{figure}

\noindent \textbf{Quantum simulation.} The Hamiltonian describing
the rotation of a linear molecule and its interaction with an external
field is given by \citep{boyd1992nonlinear,Friedrich1995Alignment,Friedrich1995Polarization}
\begin{equation}
\hat{H}(t)=B\hat{J}(\hat{J}+1)+V(t),\label{Hamiltonian}
\end{equation}
where $B=\hbar^{2}/(2I)$ is the rotational constant and the interaction
with the electric field of a non-resonant laser pulse is given by
$V(t)=-(\Delta\alpha/4)\sin^{2}\theta\left[E_{Y}(t)\cos(\phi)+E_{Z}(t)\sin(\phi)\right]^{2}$.
Here $E_{Y}$ and $E_{Z}$ are the envelopes of the field components
along the $Y$ and $Z$ axes, respectively, $\theta$ and $\phi$
are the polar and azimuthal angles of the molecular axis with respect
to the laboratory-fixed $X$ and $Y$ axes. The Hamiltonian $\hat{H}$
is expressed in the basis of eigenfunctions $|J,M\rangle$ of the
angular momentum operator $\hat{J}$, where $J$ and $M$ are the
total angular momentum and its projection on the $X$ axis. The time-dependent
Schr{\"o}dinger equation $i\hbar\partial_{t}|\Psi(t)\rangle=\hat{H}(t)|\Psi(t)\rangle$
is solved by numerical exponentiation (see Expokit \citep{sidje1998Expokit}).
In addition, the angular distribution $P_{J_{0},M_{0}}(\theta,\phi,t)=|\sum_{J,M}C_{J,M}(t)Y_{J,M}(\theta,\phi)|^{2}$
is calculated for each initial state $|\Psi(t=0)\rangle=|J_{0},M_{0}\rangle$,
where $C_{J,M}(t)=\langle J,M|\Psi(t)\rangle$ is the projection coefficient
and $Y_{J,M}(\theta,\phi)$ is the spherical harmonic function. The
final probability distribution is obtained by weighted averaging with
the Boltzmann weight.

\section{Experimental and Numerical Results \label{sec:Experimental-results}}

In our experiments, we use linear $\mathrm{N_{2}O}$ molecules, due
to their relatively long revival period $T_{\mathrm{rev}}=39.9$ ps
\citep{Hongyan2010Coherent} and the absence of $1/4$ revival. This
provides an extended observation window, of $T_{\mathrm{rev}}/2\approx20$
ps, for inducing and observing the molecular UDR echo phenomenon.
As described in the Experimental Setup (Sec. \ref{sec:Experimental-Methods}),
molecules are excited by a pair of time-delayed polarization-twisted
laser pulses which are followed by a strong circularly polarized probe
pulse, inducing Coulomb explosion. Here, the double ionization channel
$\left(\mathrm{N_{2}O}+n\hbar\omega\rightarrow\mathrm{N}^{+}+\mathrm{NO}^{+}+2e^{-}\right)$
is selected to reproduce the direction of the molecular axis based
on the relative momentum between the $\mathrm{N}^{+}$ and $\mathrm{NO}^{+}$
fragments originating from the same parent molecule. Using the described
pump-probe method, the evolution of the rotational wave packet of
$\mathrm{N_{2}O}$ is reconstructed.

\begin{figure}[!b]
\begin{centering}
\includegraphics[width=1\columnwidth]{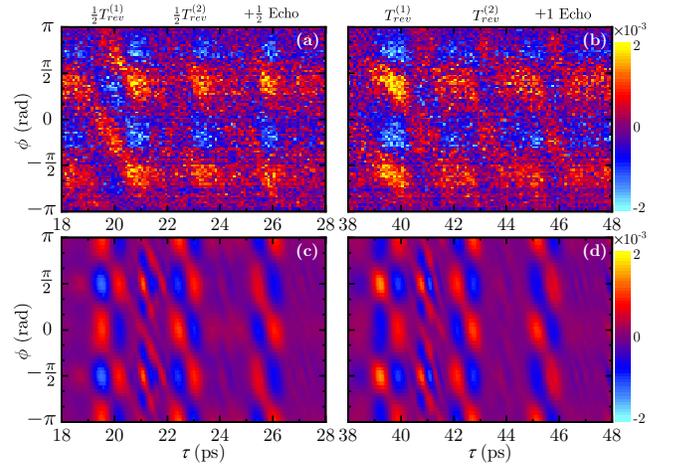}
\par\end{centering}
\caption{(a,b) Experimentally measured and (c,d) calculated (quantum mechanically)
time-dependent angular distributions for the case shown in Fig. \ref{fig:Experiment_condition1}.
\label{fig:Experiment_condition1_revival}}
\end{figure}

Figure \ref{fig:Experiment_condition1} shows the time-resolved distribution
of azimuthal angle $\phi$ in the $YZ$ plane (measured with respect
to the $Y$ axis). The distribution is normalized at each time step
and a background signal is subtracted. The background signal is obtained
by averaging the angular distribution over the range of delays between
$\tau=1.5$ ps and $\tau=2.5$ ps. As shown in Fig. \ref{fig:setup}(b),
the polarization of $\mathrm{P_{1}}$ twists clockwise ($\mathbf{e}_{f}\rightarrow\mathbf{e}_{s}$,
when observed along the negative $X$ axis) from $\pi/4$ to $-\pi/4$
with respect to the $Y$ axis, inducing molecular UDR in the plane
of polarization twisting ($YZ$ plane). Accordingly, the angular distribution
plotted in Fig. \ref{fig:Experiment_condition1}(a) exhibits the typical
features of molecular UDR, namely the probability density focuses
near $\pi/4$ $\left(-3\pi/4\right)$ and then moves towards $-\pi/4$
$\left(3\pi/4\right)$. Due to the dispersion of molecular angular
velocities, the angular distribution becomes isotropic once again
around $\tau\approx1.5$ ps. A second, weaker polarization-twisted
pulse $\mathrm{P_{2}}$, twisting in the same sense as $\mathrm{P_{1}}$,
interacts with $\mathrm{N_{2}O}$ at $\tau=3$ ps, producing an immediate
response similar to that following $\mathrm{P_{1}}$. At twice the
delay ($\tau\approx6$ ps), the molecular UDR echo appears. The echo
rotates clockwise, which can be clearly seen in the polar plots for
the consecutive probe delays $\tau=6.05,\,6.35,\,6.65\,\mathrm{ps}$
shown in Figs. \ref{fig:Experiment_condition1}(b)-\ref{fig:Experiment_condition1}(d).
Numerically calculated angular distributions (both classical and quantum)
are in good agreement with the experimental results and display similar
molecular UDR echo. The clockwise sense of molecular UDR echo is consistent
with the twisting direction of the two pulses $\mathrm{P_{1,2}}$
and is in agreement with the qualitative picture presented in Section
\ref{sec:Phase-space-analysis}.

We continue with the analysis of long-term behavior and compare it
with the results of quantum numerical simulation. As shown in Fig.
\ref{fig:Experiment_condition1_revival}, the half and full revivals
induced by the polarization-twisted pulses $\mathrm{P_{1}}$ and $\mathrm{P_{2}}$
are marked as $nT_{\mathrm{rev}}^{(1)}$ and $nT_{\mathrm{rev}}^{(2)}$
($n=1/2$ or 1), respectively. In addition, the echo caused by half
and full revivals are marked as ``$+1/2$ Echo'' and ``$+1$ Echo'',
and are visible at $\tau\approx26$ ps and $\tau\approx46$ ps, respectively.
Both the experimental and simulated results show that the molecules
coming to alignment during the revival and UDR echo rotate clockwise,
which is in the same sense as the twisting of both polarization-twisted
pulses $\mathrm{P}_{1}$ and $\mathrm{P}_{2}$ and it is similar to
the behavior observed during the first echo (at $\tau\approx6$ ps).

\begin{figure}[!t]
\begin{centering}
\includegraphics[width=1\columnwidth]{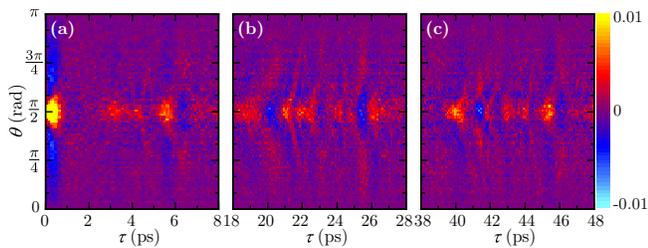}
\par\end{centering}
\caption{Experimentally measured time-dependent angular distributions of the
polar angle $\theta$ for the case shown in Fig. \ref{fig:Experiment_condition1}.
\label{fig:Experiment_condition1_theta}}
\end{figure}

Figure \ref{fig:Experiment_condition1_theta} reveals a more comprehensive
picture of the rotational dynamics. In addition to the distribution
of the azimuthal angle $\phi$ confined to the plane of polarization
twisting, we consider the time-dependent angular distribution of the
polar angle $\theta$. During the interaction with the polarization-twisted
pulses $\mathrm{P_{1,2}}$, the $\mathrm{N_{2}O}$ molecules are pulled
towards the $YZ$ plane so that the distribution of polar angle $\theta$
centers around $\theta=\pi/2$. The same behavior of the polar angle
$\theta$ is reproduced during the molecular UDR echoes, at $\tau\approx6,\,26,\,46\,\mathrm{ps}$.
To summarize, both degrees of freedom, $\theta$ and $\phi$, demonstrate
the echo phenomenon.

\begin{figure}[!t]
\begin{centering}
\includegraphics[width=7cm]{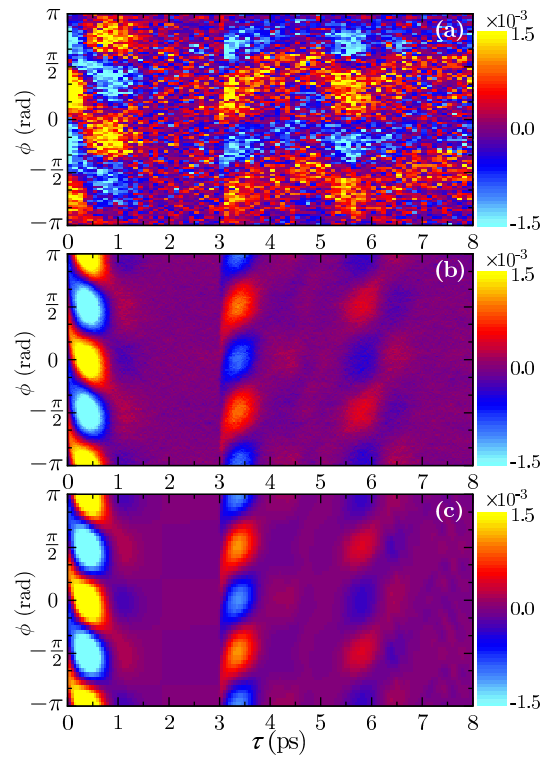}
\par\end{centering}
\caption{(a) Experimentally measured angular distribution of the angle $\phi$
as a function of time. (b) Results of the classical, and (c) quantum
simulations. The conditions used are the same as the case shown in
Fig. \ref{fig:Experiment_condition1}, except that the polarization-twisted
pulse $\mathrm{P_{2}}$ twists counterclockwise, from $\pi/4$ to
$3\pi/4$. \label{fig:Experiment_condition2}}
\end{figure}

In the first experiment (Fig. \ref{fig:Experiment_condition1}), the
twisting sense of the two polarization-twisted pulses $\mathrm{P_{1}}$
and $\mathrm{P_{2}}$ was the same, leading to a molecular UDR echo
which rotates in the same sense. In the second experiment, the sense
of polarization twisting of $\mathrm{P_{2}}$ was reversed and it
was set to twist counterclockwise from $\phi=\pi/4$ to $3\pi/4$
{[}see Fig. \ref{fig:setup}(c){]}. The angular distributions at $\tau\approx6$
ps (around the echo time) show a molecular UDR echo rotating in the
counterclockwise sense, from $\pi/4$ $(-3\pi/4)$ to $3\pi/4$ $(-\pi/4)$,
as shown in Fig. \ref{fig:Experiment_condition2}. In other words,
the rotation sense of the UDR echo follows the polarization twisting
of the second polarization-twisted pulse $\mathrm{P}_{2}$. This observation
is completely consistent with qualitative analysis presented in Sec.
\ref{sec:Phase-space-analysis}.

Finally, we would like to point out a slight difference between the
results of the classical 2D simulation presented in Fig. \ref{fig:angular_distributions_echo}(b)
and the results of the experiment and the 3D simulations presented
in Fig. \ref{fig:Experiment_condition2}. Notice that in Fig. \ref{fig:angular_distributions_echo}(b)
at the beginning of the echo event (at $t\approx5.5$ ps), the molecules
come to alignment at $\phi=0,\pi$, while in Fig. \ref{fig:Experiment_condition2}
the alignment starts at $\phi=\pm\pi/2$ (just like after the second
polarization-twisted pulse). This fine difference between the 2D model
and fully three dimensional classical/quantum simulations is discussed
in detail in Appendix \ref{App:LinearandTwistedPulses}.

\section{Conclusions \label{sec:Conclusions}}

We report the observation of molecular UDR echoes. In contrast to
previous studies on molecular alignment echoes utilizing linearly
polarized pulses, here the polarization-twisted pulses deliver a non-zero
average angular momentum to the molecular ensemble. This added angular
momentum causes a unique UDR echo dynamics, where the molecules align
during the echo, and while being aligned also rotate in unison in
a preferred sense. We observe experimentally and confirm it theoretically,
that the sense of the molecular UDR echo can be controlled by choosing
the sense of polarization twisting of the second pump pulse $\mathrm{P}_{2}$.

For the first measurement of the effect presented here, we utilized
the powerful COLTRIMS technique, which provides full spatio-temporal
information about the molecular dynamics. In principle, the molecular
UDR echoes can also be studied by robust all-optical techniques. For
example, a purely optical technique was recently demonstrated \citep{Bert2020},
allowing to detect not only the optical birefringence developing during
alignment, but also the sense of molecular UDR echo induced by the
polarization-twisted pulses.

Since their discovery, molecular rotation echoes were successfully
used in studies of molecular relaxation in condensed gases \citep{Rosenberg2018,Zhang2019,Ma2019,Rosenberg2020}.
UDR echoes may open new avenues in studying the kinetics of dense
gases of highly excited UDR molecules, which were shown to exhibit
unique collisional dynamics and non-trivial relaxation paths to thermal
equilibrium \citep{Khodorkovsky2015,Milner2015}.

\section{Acknowledgments}

\textcolor{black}{This work is supported by the National Key R\&D
Program of China (Grant No. 2018YFA0306303), the National Natural
Science Foundation of China (Grant Nos. 11834004, 61690224, 11621404
and 11761141004), the 111 Project of China (Grant No. B12024), the
Israel Science Foundation (Grant No. 746/15), the ICORE program ``Circle
of Light'', and the ISF-NSFC (Grant No. 2520/17).} I.A. acknowledges
support as the Patricia Elman Bildner Professorial Chair. This research
was made possible in part by the historic generosity of the Harold
Perlman Family.

\appendix

\section{Comparing groups of clockwise and counterclockwise rotating molecules
\label{App:Population difference}}

\begin{figure}[!bhtp]
\begin{centering}
\includegraphics[width=1\columnwidth]{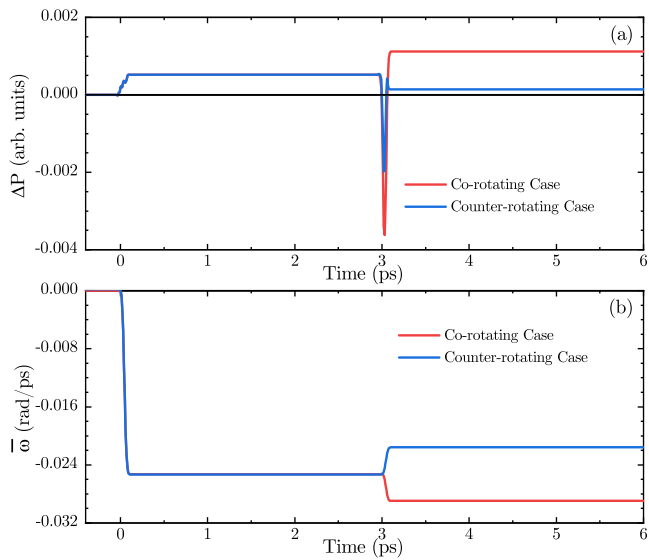}
\par\end{centering}
\caption{(a) Time-dependent population difference between molecules with negative
and positive angular velocities, $\triangle P=\int_{-\infty}^{0}p(\omega)\mathrm{d}\omega-\int_{0}^{\infty}p(\omega)\mathrm{d}\omega$
and (b) average angular velocity, $\overline{\omega}=\int\omega p(\omega)\mathrm{d}\omega$
for the two cases shown in Fig. \ref{fig:angular_distributions_echo}.
\label{fig:Population}}
\end{figure}

Figure \ref{fig:Population} shows the population difference and the
average angular velocity of clockwise and counterclockwise rotating
molecules calculated using the 2D model {[}Eq. (\ref{eq:2Dmodel}){]}.
Before the pulses, the distribution of angular velocities is symmetric.
As a result, the populations of molecules with negative and positive
angular velocities are equal, and the average angular velocity is
zero. The polarization direction of the first pulse twists clockwise,
leading to a population difference between molecules with negative
and positive angular velocities $\triangle P=0.052\%$, as well as
non-zero average angular velocity, $\overline{\omega}=-0.0253$ rad/ps.
When the second pulse also twists clockwise, the population difference
($\triangle P=0.112\%$) and the absolute value of average angular
velocity ($\overline{\omega}=-0.0290$ rad/ps) are enhanced. On the
other hand, when the second pulse twists in the opposite sense (counterclockwise),
the population difference ($\triangle P=0.014\%$) and the average
angular velocity ($\overline{\omega}=-0.0216$ rad/ps) keep the same
sign, as after the first pulse (absolute values are decreased). Therefore,
in both cases, the molecules rotate on average (including all molecules)
clockwise after the application of the two pulses.

\begin{figure*}[t]
\begin{centering}
\includegraphics[width=1.5\columnwidth]{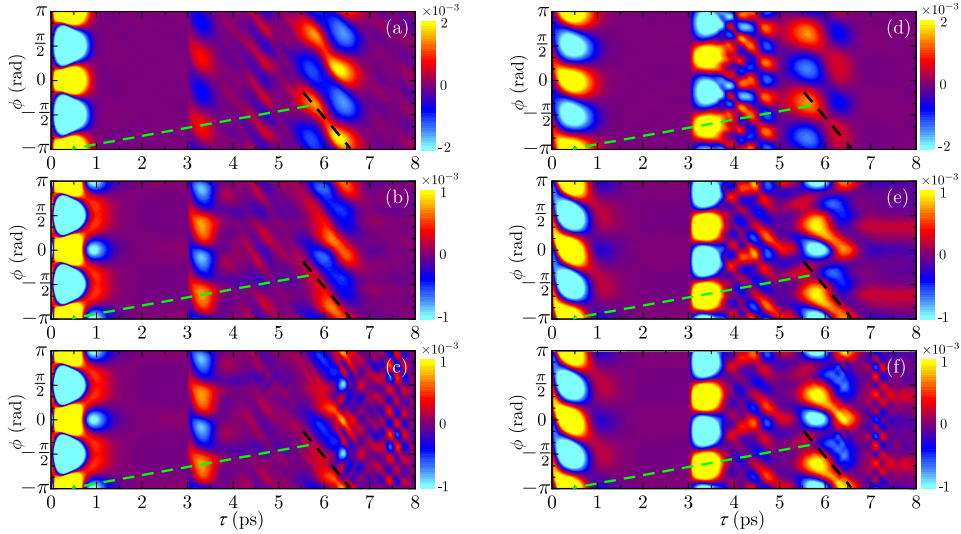}
\par\end{centering}
\caption{On the left: time-dependent angular distributions calculated using
(a) the 2D model {[}Eq. (\ref{eq:2Dmodel}){]}, (b) fully three-dimensional
classical, and (c) quantum simulations for the case when a linearly
polarized pulse is applied first ($\mathrm{P_{1}}$), and it is followed
by a polarization-twisted pulse ($\mathrm{P_{2}}$). The pulse parameters
are the same as the case shown in Fig. \ref{fig:Experiment_condition1}.
Linearly polarized pulse $\mathrm{P_{1}}$ is polarized along the
$Y$ axis and $\mathrm{P_{2}}$ twists clockwise from $7\pi/12$ to
$\pi/12$. On the right: Time-dependent angular distributions calculated
using (d) the 2D model {[}Eq. (\ref{eq:2Dmodel}){]}, (e) fully three-dimensional
classical, and (f) quantum simulations for the case when a polarization-twisted
pulse is applied first ($\mathrm{P_{1}}$), and it is followed by
a linearly polarized pulse ($\mathrm{P_{2}}$). $\mathrm{P_{1}}$
twists clockwise from $\pi/4$ to $-\pi/4$, while $\mathrm{P_{2}}$
is polarized along $\pi/3$ with respect to the $Y$ axis. \label{fig:LinearandTwistedPulses}}
\end{figure*}

\section{Analysis of the slight difference between the results of 2D and 3D
simulations \label{App:LinearandTwistedPulses}}

In this appendix, we consider more closely the difference between
the results of the classical 2D simulation presented in Fig. \ref{fig:angular_distributions_echo}(b)
and the results of the experiment and 3D simulations presented in
Fig. \ref{fig:Experiment_condition2}. The molecular alignment during
the UDR echo obtained using the 2D model is rotated relative to the
alignment during the application of the second polarization-twisted
pulse. Similar ``rotated echoes'' were observed in \citep{Lin2017},
where two linearly polarized pump pulses were used to induce alignment
echoes. The pulses were polarized at different angles and the alignment
echo appeared rotated by twice the relative angle between the pulses.
To check this and further analyze the difference between the results
of 2D and 3D simulations, we substitute one of the polarization-twisted
pulses by a linearly polarized pulse.

Figure \ref{fig:LinearandTwistedPulses} shows the angular distributions
obtained using the 2D model {[}Eq. (\ref{eq:2Dmodel}){]}, fully three
dimensional classical, and quantum simulations. Results for two cases
are presented: (i) On the left side--the first pulse is linearly
polarized along the $Y$ axis, while the second pulse is polarization-twisted,
and (ii) On the right side--the first pulse is polarization-twisted,
while the second pulse is linearly polarized at an angle to the $Y$
axis. The pulses' parameters are the same as in the experiments (see
Fig. \ref{fig:Experiment_condition1}). In both cases, molecular UDR
echo appears at twice the delay between the pulses, and it is qualitatively
similar. In addition, regardless of whether the polarization-twisted
pulse is applied first or second, the molecular UDR echo rotates in
the same sense as the polarization twisting.

The results obtained using the 2D model once again differ from the
results of 3D simulations. Figs. \ref{fig:LinearandTwistedPulses}(a)
and \ref{fig:LinearandTwistedPulses}(d) show that the location of
the initial alignment during the UDR echo is rotated by an additional
angle when compared to the results of fully 3D classical and quantum
simulations. To summarize, although the 2D model predicts the existence
of rotated echo, this prediction is not verified in the more accurate
3D simulations/experiment. Therefore, we conclude that when at least
one of the pulses used to induce the UDR echo is polarization-twisted,
the echo is not ``rotated''.

\end{document}